\documentclass[prd, aps, superscriptaddress, preprintnumbers, twocolumn, floatfix, nofootinbib]{revtex4}
\pdfoutput=1

\usepackage{amsfonts}
\usepackage{amsmath}
\usepackage{amssymb}
\usepackage{bm}
\usepackage{dcolumn}
\usepackage{graphicx}   
\usepackage[latin1]{inputenc}
\usepackage{latexsym}
\usepackage{rotating}
\usepackage{hyperref}
\usepackage{graphicx}
\usepackage{color}

\newcommand\be{\begin{equation}}
\newcommand\ba{\begin{eqnarray}}
\newcommand\ee{\end{equation}}
\newcommand\ea{\end{eqnarray}}

\newcommand{\cHm}{{{\cal{H}_{-}}}}
\newcommand{\cHp}{{{\cal{H}_{+}}}}

\begin{document}

\title{Ekpyrotic Cosmology with a Zero-Shear S-Brane}

\author{Robert Brandenberger}
\email{rhb@physics.mcgill.ca}
\affiliation{Department of Physics, McGill University, Montr\'{e}al, QC, H3A 2T8, Canada}

\author{Ziwei Wang}
\email{ziwei.wang@mail.mcgill.ca}
\affiliation{Department of Physics, McGill University, Montr\'{e}al, QC, H3A 2T8, Canada}

\date{\today}

\begin{abstract}

In a recent paper \cite{us} we proposed a mechanism for a continuous transition between a contracting Ekpyrotic phase and the Standard Big Bang phase of expansion: the bounce is generated by an S-brane which represents the effects of higher mass string states in the low energy effective field theory. We showed that gravitational waves on cosmological scales obtain a nearly scale-invariant spectrum. Here, we study the cosmological fluctuations in this setup, assuming that the S-brane has zero shear. We find a nearly scale-invariant spectrum of cosmological perturbations with a slight red tilt. The scenario yields two consistency relations for cosmological observations, the first one relating the tensor to scalar ratio with the scalar spectral tilt, the second relating the tensor tilt to the scalar tilt. The predicted tensor to scalar ratio is within the reach of upcoming CMB observations. The tensor tilt is blue.

\end{abstract}

\pacs{98.80.Cq}
\maketitle

\section{Introduction} 
\label{sec:intro}

In light of the recent challenges for inflationary cosmology (both the inability to embed canonical single field slow roll inflation into string theory as a consequence of the {\it swampland criteria} (see \cite{orig} for original articles and \cite{swamprev} for reviews) and the {\it Trans-Planckian Censorship Conjecture} \cite{BeVa, BBLV}) it is of great interest to explore possible alternatives to inflation and further develop their predictions for cosmological observations.

The Ekpyrotic scenario \cite{Ekp} is a promising alternative to cosmological inflation for explaining the isotropy of the microwave background and the spatial flatness of today's universe. The scenario is based on the assumption that the universe begins in a phase of slow contraction. In the context of Einstein gravity such a slow contracting phase can be generated by a scalar matter field $\varphi$ with a negative exponential potential with the exponent chosen such that the homogeneous scalar field trajectory has an equation of state paramter $w \gg 1$, where $w$ is the ratio between the pressure and energy density, and the scale factor evolves as
\be
a(t) \, \sim \, (-t)^p \,\,\,\, (t < 0) \, 
\ee
with \footnote{The Pre-Big-Bang scenario of \cite{PBB} is the case $p = 1$ and can be obtained in the context of dilaton gravity.} 
\be
p \, \ll \, 1 \, . 
\ee
In terms of conformal time $\tau$ the scale factor evolves as
\be
a(\tau) \, \sim \, (- \tau)^q \,\,\,\, (\tau < 0)
\ee
with
\be
q \, = \, \frac{p}{1 - p} \, .
\ee
A nice feature of the Ekpyrotic scenario is the fact that anisotropies and initial inhomogeneities are diluted in the contracting phase. Thus, the homogeneous and isotropic contracting phase space trajectory is an attractor in initial condition space \cite{Erickson}.

The main challenge for the Ekpyrotic scenario has been to obtain a well-controlled transition from the contracting Ekpyrotic phase to the expanding phase which is postulated to evolve as in Standard Big Bang cosmology, i.e. beginning with a radiation-dominated phase after the transition \footnote{In the original paper \cite{Ekp} it was argued that the transition would be singular from the point of view of an effective field theory coupled to Einstein gravity. More recently, however, there have been a lot of attempts to realize the Ekpyrotic scenario in the context of a non-singular effective field theory \cite{Ekp-NS}. Note that violations of the Null Energy Condition are required in order to obtain a non-singular bounce.} . A related challenge has been to obtain a robust computation of the spectrum of cosmological perturbations and gravitational waves, assuming that these fluctuations begin early in the contracting phase in their vacuum state.

In a recent paper, we \cite{us} have proposed a mechanism which yields a nonsingular transition between contraction and expansion. We argued that, in the context of string theory, at the string density new stringy degrees of freedom must be included in the effective action for the background and fluctuations. These terms yield an {\it S-brane}, a space-like hypersurface which is reached when the background density hits the string density. Since such an S-brane has vanishing energy density and negative pressure, it yields the violation of the energy conditions required to obtain a transition between contraction and expansion. 

The second main result of \cite{us} was the demonstration that, on super-Hubble scales, gravitational modes passing through the S-brane experience an enhancement of their amplitude proportional to $k^{-1}$, where $k$ is the comoving wavenumber. This amplification is precisely the correct one required to turn an initial vacuum spectrum into a scale-invariant one. The same mechanism also changes the spectrum of the canonical curvature fluctuation variable $v$, but the amplitude of the enhancement was found to be smaller than the amplitude of the enhancement of the gravitational waves. It this were the only amplification mechanism of the scalar spectrum, an unacceptably large tensor to scalar ratio would result.

However, the evolution of the scalar metric fluctuations is more complicated than that of the tensor modes. The precise specification of the location of the S-brane is crucial. Here, we demontrate that, if the S-brane has no shear, then the growing mode of the scalar field fluctuation $\delta \varphi$ seeds an approximately scale-invariant spectrum of curvature fluctuations in the expanding phase. Combined with the results of \cite{us} for the gravitational wave spectrum we obtain two consistency relations for current observables, the first relating the tensor to scalar ratio $r$ with the scalar tilt $n_s$, the second relating the tensor tilt $n_t$ to the scalar tilt. The first relation is
\be \label{cons1}
r \, = \, {\cal B} (1 - n_s)^2 \, ,
\ee
where ${\cal B}$ is a constant of the order one, and the second relation is
\be \label{cons2}
n_t \, = \, (1 - n_s) \, .
\ee
Since the tilt of the scalar spectrum is red, (\ref{cons2}) represents a blue tilt of the tensor spectrum, like \cite{BNPV} in the case of String Gas Cosmology \cite{BV, SGrevs}.

In the folliowing section we discuss the computation of scalar metric fluctuations in Ekpyrotic cosmology, reviewing known results and emphasizing the role of the specific location of the matching surface between the contracting and expanding phases. In Section 3 we review the S-brane scenario introduced in \cite{us}. Section 4 provides a summary of our calculations, and in Section 5 we discuss some implications of our results. We assume a spatially flat homogeneous and isotropic background space-time with scale factor $a(t)$, where $t$ is physical time. It is often convenient to use conformal time $\tau$ determined by $dt = a(t) d\tau$. The Hubble expansion rate is $H \equiv {\dot{a}}/a$, and its inverse is the Hubble radius, the length scale which plays a key role in the evolution of cosmological fluctuations. We use units in which the speed of light and Planck's constant are set to $1$. Comoving spatial coordinates are denoted by ${\bf{x}}$, and the corresponding comoving momentum vector is ${\bf{k}}$ (its magnitude is written as $k$). The reduced Planck mass is denoted by $m_{pl}$ and it is related to Newton's gravitational constant $G$ via $m_{pl}^{-2} = 8 \pi G$.

\section{Curvature Fluctuations in Ekpyrotic Cosmology}

During a phase of Ekpyrotic contraction ($\tau < 0$), fluctuations of the scalar field $\varphi$ which start as vacuum perturbations in the far past acquire \cite{Ekp} a nearly scale-invariant power spectrum ${\cal{P}}_{\delta \varphi}(k)$ 
\be
{\cal{P}}_{\delta \varphi}(k) \, \equiv \, \frac{1}{2 \pi^2} k^3 |\delta \varphi_k|^2 \, \sim \, k^{n_s - 1} \, 
\ee
(where $\delta \varphi_k$ is the Fourier mode of $\delta \varphi$) with a red tilt given by the spectral index
\be
n_s - 1 \, = \, - 2q \, ,
\ee
to leading order in $q$, while the spectrum of gravitational waves (with the same initial conditions) retains its vacuum form, modulo a small blue tilt given by
\be
\delta n_t \, = \, 2q \, ,
\ee
again to leading order in $q$. Note that the tensor tilt of the canonically normalized variable 
\be
{\tilde{h}} \, \equiv \, a h \, ,
\ee
(where $h$ is the amplitude of a particular gravitational wave polarization mode) is defined via (we are using the conventions from \cite{Baumann})
\be
{{\cal P}}_{{\tilde{h}}}(k) \, \equiv \, \frac{1}{2 \pi^2} |{\tilde{h}}_k|^2 \, \sim \, k^{n_t} \, 
\ee
(${\tilde{h}}_k$ again denoting the Fourier mode of ${\tilde{h}}$). The transformation of the spectrum of $\delta \varphi$ is a consequence of the fact that $\delta \varphi_k$ has a mode which grows on super-Hubble scales, whereas ${\tilde{h}}_k$ does not have such a growing mode.
 
On the other hand, as pointed out in \cite{Lyth, FB}, the canonical fluctuation variable $v$ \cite{Sasaki, Mukh} which describes the amplitude of curvature fluctuations in comoving coordinates, does not grow on super-Hubble scales. In fact, the dominant mode of $v_k$ slightly decreases, like the leading mode of ${\tilde{h}}_k$. This implies that the spectrum of fluctuations of the comoving gauge curvature at the end of the contracting phase remains close to vacuum. In fact, it acquires the same slight blue tilt which the spectrum of gravitational waves does.

The fact that the spectra of $\delta \varphi$ and $v$ have a very different spectral index in the contracting phase of the Ekpyrotic scenario (they have the same index in the case of single field slow-roll inflation) implies that great care must be taken to correctly evolve the fluctuations from the end of the contracting phase to the beginning of the expanding phase. In the following we will review this issue. Note that this ambiguity does not effect the evolution of the gravitational wave spectrum since gravitational waves are gauge-independent.

Following the notation of \cite{MFB} (see also \cite{RHBfluctsrev} for an overview) we will work in longitudinal gauge, a coordinate system in which the metric (including scalar fluctuations) is diagonal and given by the line element
\be \label{pertmetric}
ds^2 \, = \, a(\tau)^2 \bigl[ (1 + 2 \Phi) d\tau^2 - (1 - 2 \Phi) d{\bf{x}}^2 \bigr] \, ,
\ee
where $\Phi({\bf{x}}, \tau)$ is the scalar fluctuations variable, the relativistic generalization of the Newtonian gravitational potential, and we have assumed the absence of anisotropic matter stress, an assumption which is satisfied at linear order in the case of a scalar matter field. The scalar matter field $\varphi$ can be decomposed into its background value $\varphi_0(\tau)$ and the fluctuation $\delta \varphi({\bf{x}}, \tau)$:
\be \label{pertmatter}
\varphi({\bf{x}}, \tau) \, = \, \varphi_0(\tau) + \delta \varphi({\bf{x}}, \tau) \, .
\ee

The amplitude of the late time anisotropies in the microwave background and of the inhomogeneities which lead to the observed large-scale structure of the universe are determined by the spectrum of the curvature fluctuation variable ${\cal{R}}$, the curvature perturbation in comoving coordinates. This quantity is closely related to the Sasaki-Mukhanov variable $v$ \cite{Sasaki, Mukh},  the variable in terms of which the action for linear cosmological perturbations has a canonical kinetic term. This variable is
\be
v \, = \, a \bigl( \delta \varphi + \frac{{\dot{\varphi_0}}}{H} \Phi \bigr) \, ,
\ee
and its Fourier mode obeys the equation
\be \label{veq}
v_k^{\prime \prime} + \bigl( k^2 - \frac{z^{\prime \prime}}{z} \bigr) v_k \, = \, 0 \, ,
\ee
where
\be
z \, \equiv \, a \frac{\varphi_0^{\prime}}{{\cal{H}}} \, ,
\ee
with a prime denoting a derivative with respect to conformal time and ${\cal{H}} \equiv a^{\prime}/a$.

In the phase of Ekpyrotic contraction, $z$ and $a$ differ only by a constant, and hence (on super-Hubble scales where the $k^2$ term is negligible) (\ref{veq}) becomes
\be
v_k^{\prime \prime} + q(1 - q) v_k \, = \, 0 \, ,
\ee
with a dominant solution
\be
v_k \, \propto \, (-\tau)^{q(1 - q)} \, \sim \, (-\tau)^q \, ,
\ee
a solution which is in fact slightly decaying as a function of time. As a consequence, an initial vacuum spectrum for fluctuations of $v$ does not get transformed into a scale-invariant one. Rather, the spectrum acquires a slight blue tilt $\Delta n_s > 0$ compared to a vacuum spectrum:
\be
\Delta n_s \, = \, 2q \, ,
\ee
to leading order in the small parameter $q$. To derive this relation, note that $v_k$ oscillates with constant amplitude on sub-Hubble scales, and then scales as $(-\tau)^q$ on super-Hubble scales. Hence, the total change in amplitude of the mode $v_k$ compared to its vacuum value is
\be
\frac{v_k(\tau)}{v_k(\tau_H(k))} \, \sim \, (-\tau_H(k))^{-q} \, \sim \, k^q \, ,
\ee
where $\tau_H(k)$ is the time when the mode $k$ exits the Hubble radius (see Figure 1). Note that the canonical variable ${\tilde{h}}$ for gravitational waves obeys the same equation (\ref{veq}).

A different view of the evolution of scalar metric fluctuations in the phase of Ekpyrotic contraction is obtained by following the relativistic potential $\Phi$, or equivalently the rescaled variable 
\be \label{udef}
u \, \equiv \, \frac{m_{pl}}{{\cal{H}}} a \Phi \, .
\ee
It obeys the mode equation \cite{KOST2, DV}
\be \label{ueq}
u_k^{\prime \prime} + \bigl( k^2 - \frac{2}{a} {\cal{H}}^2 - \frac{a^{\prime \prime}}{a} \bigr) u_k \, = \, 0 \, ,
\ee
which on super-Hubble scales (and in the case of Ekpyrotic contraction) becomes
\be
u_k^{\prime \prime} - q (-\tau)^{-2} u_k \, = \, 0 \, ,
\ee
which leads to a growing mode with
\be \label{ugrowing}
u_k \, \sim \, (-\tau)^{-q} \, \sim \, a(\tau)^{-1} \, ,
\ee
and a decaying mode with
\be \label{udecaying}
u_k \, \sim \, (-\tau)^{1 + q} \, ,
\ee
always to leading order in $q$. This implies that in the infrared limit there are two modes for $\Phi$, a growing mode with
\be \label{growing}
\Phi_k(\tau) \, = \, A(k) \frac{{\cal{H}}}{a}(\tau) \frac{a(\tau_H(k))}{a(\tau)} \, ,
\ee
and a constant mode
\be \label{decaying}
\Phi_k(\tau) \, = \, B(k) \, ,
\ee
where the coefficients $A(k)$ and $B(k)$ are determined by the vacuum initial conditions which are \footnote{Note that the factor $k^{-3/2}$ comes from the fact that $u$ is related to the canonical fluctuation variable by a factor which in the ultraviolet scales as $k^{-1}$, and the canonical variable having vacuum initial conditions $\sim k^{-1/2}$.}
\be \label{vacuum}
{\rm{lim}}_{\tau \rightarrow - \infty} u_k \, = \, \frac{1}{k^{3/2}} e^{-ik\tau} \, .
\ee
It is the factor of $k^{-3/2}$ in the vacuum normalization of $u$ which leads to the scale-invariance of the power spectrum of $u$. In fact, the slight growth of $u$ on super-Hubble scales yields a red tilt of the spectrum with
\be
n_s - 1 \, = \, - 2q \, ,
\ee
which should be compared to the tensor tilt of
\be
n_t \, = \, 2q \, .
\ee

Since at the end of the contracting phase the dominant modes of $\Phi$ and $v$ have very different scalings, the key question is which of the variables is continuous when passing through the S-brane. In the case of single field slow-roll inflation the canonical variable $v$ is continous while $\Phi$ jumps if we treat reheating as instantaneous (see e.g. \cite{BK}). In the case of bouncing cosmologies with a smooth transition from contracting to expansion (which always requires new physics input), one can often (see e.g. \cite{bounce} for some specific examples, and \cite{BP} for a review) show that it is again the variable $v$ which determines the spectrum of the dominant mode at late times (however, an example where this is not true was studied in \cite{Xue}). In our case, the transition is not smooth and happens on a space-like hypersurface, the S-brane. The transfer of fluctuations across a space-like matching surface was studied in detail in \cite{DV} (see also \cite{DM}), generalizing the Israel matching conditions \cite{Israel}. It was found that the location of the surface in space-time is crucial in order to determine whether $\Phi$ or $v$ determine the spectrum of late time fluctuations. It was found that if the hypersurface is at a constant density surface (as assumed in \cite{us}), then the dominant mode of $\Phi$ in the contracting phase does not couple to the dominant mode in the expanding phase, while $v$ is continuous, and it is hence the spectrum of $v$ which determines the amplitude of the late time curvature fluctuations. On the other hand, for any other matching surface, in particular in the case of a shear-free surface which we will be using in this paper, then the growing mode of $\Phi$ in the contracting phase couples to the dominant mode of the late time curvature fluctuation variable.

\section{S-Brane and Nonsingular Bounce}

We now briefly review the scenario introduced in \cite{us}. It is based on the assumption that our Ekpyrotic scenario is embedded in superstring theory. In this context, the scalar field $\varphi$ is the lightest modulus field of the particular string theory background being considered (e.g. the radius of the large extra dimension in Horava-Witten theory \cite{HW}, the original setting of the Ekpyrotic scenario \cite{Ekp}). However, as the background energy density increases, the mass of a large tower of string states no longer becomes negligible. At the string scale, this tower of states has to be included. The suggestion of \cite{us} (see also \cite{Kounnas} for a similar proposal in a slightly different setting) is that these degrees of freedom must be included as an {\it S-brane}, a relativistic space-like hypersurface which is located where the density hits the string scale $\eta_s$. Specifically, the low energy action of the theory is assumed to take the form
\ba \label{action}
S \, &=& \, \int d^4x \sqrt{-g} \bigl[ R + \frac{1}{2} \partial_{\mu} \varphi \partial^{\mu} \varphi
- V(\varphi) \bigr] \nonumber \\
& & - \int d^4x \kappa \delta(\tau - \tau_B) \sqrt{\gamma} \, ,
\ea
where $\tau_B$ is the time when the string density is reached, $R$ is the Ricci scalar of the four-dimensional space-time metric $g_{\mu \nu}$ with determinant $g$, $\varphi$ is our Ekpyrotic scalar field, $\gamma_{ij}$ is the induced metric on the hypersurface with determiant $\gamma$, and $\kappa$ is the tension of the S-brane. The tension of the brane is set by the string scale $\eta_s$
\be \label{kappa}
\kappa \, \equiv \, N \eta_s^3 \, ,
\ee
where $N$ is a large integer given by the number of new states which become low mass.

Since the S-brane is a space-like relativistic object, its energy density vanishes and it has negative pressure. This can be understood in analogy with the energy-momentum tensor of a domain wall which has vanishing pressure in orthogonal direction and negative pressure (i.e. positive tension) in the parallel directions. Hence, the S-brane leads to a violation of the null energy condition, such a violation being required to obtain a transition from contraction to expansion. If we denote the background time at the position of the S-brane by $t_B$, then we can integrate the Friedmann equation across the brane and obtain a change
\be \label{change1}
\delta H \,  \equiv  \,  \lim_{\epsilon \rightarrow 0} H(t_B + \epsilon) - H(t_B - \epsilon) \, = 
\,  4 \pi G \kappa \, .
\ee

We see that the integer $N$ has to be sufficiently large in order to allow for a transition from contraction to expansion. We will assume that the bounce is symmetric in the sense
\be
\delta H \, = \, 2 |H_{-}| \, ,
\ee
where $H_{-}$ is the value of $H$ at the end of the contracting phase when the energy density reaches the string scale. Making use of the Friedmann equation
\be \label{change2}
\delta H \, = \, \frac{2}{\sqrt{3}} \eta_s^2 m_{pl}^{-1} \, ,
\ee
comparing (\ref{change2}) with (\ref{change1}) and inserting (\ref{kappa}) leads to the condition
\be \label{Nvalue}
N \, = \, \frac{4}{\sqrt{3}} \frac{m_{pl}}{\eta_s} \, .
\ee

Our model has two free parameters, firstly the ratio of the string scale to the Planck mass, and secondly the index $q$. As we will see in the next section, the amplitude of the scalar spectrum is set by the ratio of the string scale to the Planck mass, and also depends on the value of $q$. The amplitude of the tensor modes depends on $N$, as shown in \cite{us}. In this way, two consistency relations for observables result, the first being the relation (\ref{cons2}) between the tensor and scalar tilts, and the second a relation between the tensor to scalar ratio and the scalar tilt which will be derived below.

\section{Curvature Fluctuations with a Zero Shear S-Brane}

In our previous paper we took the S-brane to be located at the surface of constant density. In this case, as shown in \cite{DV}, the growing mode of $\Phi$ in the contracting phase does not couple to the dominant mode in the expanding phase. From the point of view of the dynamics of the S-brane, it is, however, natural to assume that the brane has no shear. In this case, the brane is located at the surface of constant time in the longitudinal gauge which we are using.

The matching conditions of both the background and the cosmological fluctuations were studied in detail in \cite{DV}, and we will make use of the results derived there. The matching conditions state that firstly the induced metric computed from both sides of the brane is identical, and the second condition states that the extrinsic curvature jumps by an amount which is set by the surface tension. For the background, it is this second condition which is equivalent to the condition (\ref{change1}) which we have used above.

Scales we are interested in are in the far infrared compared to the Hubble radius. Hence, we can focus on the infrared limit of the solutions of the equation of motion for $u$. In this limit, and following the notation of \cite{DV}, we write the solution of $\Phi$ in the contracting phase as
\be
\Phi_{-}(k, \tau) \, = \, A_{-}(k) \frac{{\cal{H}}}{a^2} + B_{-}(k) \, ,
\ee
where we have absorbed the factor of $a(\tau_H(k))$ from (\ref{growing}) into the coefficient $A_{-}$. Analogously, in the expanding phase the solution can be written as
\be
\Phi_{+}(k, \tau) \, = \, A_{+}(k) \frac{{\cal{H}}}{a^2} + B_{+}(k) \, .
\ee
For a zero shear brane and neglecting fluctuations in the S-brane tension, matching the induced metric on both sides of the brane, and having the jump in the extrinsic curvature being given by the surface tension, leads to the following relations between the $A$ and $B$ coefficients (see \cite{DV})
\ba \label{matching}
A_{+} \, &=& \frac{{\cal{H}}_{-}}{{\cal{H}}_{+}} A_{-} + \frac{a_B^2}{{\cal{H}}_{+}} (B_{-} - B_{+}) \nonumber \\
B_{+} \, &=& \bigl( \frac{\cHp (\cHm^{\prime} / \cHm - \cHm) - \cHp^{\prime} + \cHp^2}{2\cHp^2 - \cHp^{\prime}} \bigr) \frac{\cHm}{a_B^2} A_{-} \nonumber \\
& & + \bigl( 1 + \frac{\cHm\cHp - \cHp^2}{2\cHp^2 - \cHp^{\prime}} \bigr) B_{-} \, ,
\ea
where $a_B$ is the value of $a$ at the transition surface \footnote{We assume that the perturbations in the surface tension are negligible in the infrared limit.}. As is evident from (\ref{matching}), the growing mode of $\Phi$ in the contracting phase couples to the dominant mode in the expanding phase. The effect of the subdominant mode in the contracting phase on $B_{+}$ is negligible. Making use of the scaling of $\cal{H}$ in the contracting phase we obtain
\be \label{matching2}
B_{+}(k) \, \simeq \, - \frac{\cHp}{a_B^2} \frac{1}{3q} A_{-}(k)  \, .
\ee

The amplitude $A_{-}(k)$ is obtained by matching to the vacuum initial conditions of (\ref{vacuum}) and yields \cite{DV}
\be
A_{-}(k) \, \simeq \, 2^{\mu} \Gamma(\mu) m_{pl}^{-1} k^{-3/2} (k \tau_B)^{-q} \, ,
\ee
where $\tau_B$ is the conformal time when the S-brane arises, $\Gamma$ stands for the gamma function, and $\mu = q + 1/2$. Making use of (\ref{matching2}) we find the following result for the late time power spectrum of $\Phi$:
\be
{\cal{P}}_{\Phi}(k) \, \simeq \, \frac{1}{2 \pi^2} (k \tau_B)^{-2q}  \bigl( \frac{\cHp}{a_B^2 m_{pl}} \bigr)^2 \frac{1}{9q^2} 2^{2\mu} \Gamma(\mu)^2 \, .
\ee
Making use of the Friedmann equation to solve for $\cHp$ we obtain
\be \label{scalarspectrum}
{\cal{P}}_{\Phi}(k) \, \simeq \, \frac{1}{2 \pi^2} (k \tau_B)^{-2q}  \bigl( \frac{\eta_s}{m_{pl}} \bigr)^4 \frac{1}{27q^2} 2^{2\mu} \Gamma(\mu)^2 \, .
\ee
We see that the spectrum is approximately scale-invariant with a small red tilt of magnitude $2q$.

If we demand that the power spectrum of $\Phi$ has the observed valuie $10^{-9}$ \cite{Planck} at the pivot scale $k_C$ (the scale of the CMB quadrupole), we find 
\be \label{stringscale}
\bigl( \frac{\eta_s}{m_{pl}} \bigr)^4 \, \simeq \, (2 \pi^2) 27 q^2 2^{-2\mu} \Gamma(\mu)^{-2} (k_C \tau_B)^{2q} 10^{-9} \, .
\ee
The amplitude of the power spectrum of gravitational waves obtained from \cite{us} is
\be \label{GWspectrum}
{\cal{P}}_{h}(k) \, \simeq \, \frac{1}{2 \pi^2} \kappa^2 m_{pl}^{-6} (k \tau_B)^{2q} \, .
\ee
Hence, the predicted tensor to scalar ratio $r$ (evaluated at the pivot scale $k_C$) is
\ba \label{Ratio}
r \, &\equiv& \, \frac{{\cal{P}}_h(k)}{{\cal{P}}_{\Phi}(k)} \\
& \simeq & \, 144 (k_C \tau_B)^{4q}  2^{-2\mu} \Gamma(\mu)^{-2} q^2
\nonumber 
\ea
which is obtained by combining (\ref{GWspectrum}) and (\ref{scalarspectrum}) and making use of the expression (\ref{Nvalue}) for $N$. 

Since the value of $q$ is given by the scalar tilt $q = (1 - n_s)/2$, (\ref{Ratio}) yields a consistency relation between the tensor to scalar ratio and the scalar tilt. Its approximate form is
\be
r \, \sim \, 36 (k_C \tau_B)^{4q} (1 - n_s)^2  \, .
\ee

\section{Conclusions and Discussion} \label{conclusion}

We have extended our work \cite{us} studying the spectrum of cosmological fluctuations in an Ekpyrotic model in which the transition from contraction to expansion is mediated by an S-brane which represents the effects of higher energy string states which enter the effective action at the string scale. In a previous paper we discovered that the coupling of the gravitational fluctuations to the S-brane leads to the conversion of an initial vacuum spectrum to a scale-invariant one. This conversion occurs both for gravitational waves and also for the Sasaki-Mukhanov variable $v$ representing the cosmological perturbations. However, this mechanism gives a greater boost to the gravitational waves then to the variable $v$. As emphasized in \cite{KOST2, DV}, in the case of a transition from contraction to expansion the variable $v$ does not necessarily carry the full information about the scalar metric fluctuations. The Newtonian gravitational potential $\Phi$ grows in the contracting phase and obtains a roughly scale-invariant spectrum. Here, we have shown that if the S-brane has zero shear (instead of being located at the constant density surface), then the dominant fluctuation mode at late time is strongly coupled to the growing mode in the contracting phase. Hence, the spectrum of cosmological perturbations at late times is approximately scale-invariant and obtains an amplitude which is larger than that of the gravitational wave spectrum.

Our model has two free parameters: the ratio between the string scale and the Planck scale, and the index $q$ which describes the rate of Ekpyrotic contraction. If we fix these two parameters in terms of the observed amplitude and tilt of the spectrum of cosmological perturbations, we obtain predictions for the tensor to scale ratio $r$, and for the tensor tilt:
\ba
n_t \, &=& \, 1 - n_s \\
r \, &=& {\cal B} (1 - n_s)^2  \, , \nonumber 
\ea
where ${\cal B}$ is a number of order one. Note that these consistency relations are different from those of canonical single field inflation (which always predicts a red tilt of the tensor spectrum). The relation between the tilts $n_t$ and $n_s$ is the same as that obtained in String Gas Cosmology, but the predicted value of $r$ has a different dependence on the parameters than what is obtained in String Gas Cosmology \cite{NBV, BNP}.

An important open problem for our scenario is to study the generation of radiation during the transition through the S-brane. Work on this topic is in progress. It would also be interesting to study the amplitude and shape of the non-Gaussianities. In inflationary cosmology the amplitude of the non-Gaussianities is suppressed by the slow-roll parameter $\epsilon$. In the Ekpyrotic scenario the analog of the inflationary slow-roll parameter is a large number, and hence one suppression mechanism of non-Gaussianities disappears. In the {\it New Ekpyrotic Scenario} \cite{New-Ekp, Ekp-NS}, a two field model in which scale-invariant fluctuations of a second scalar field $\chi$ source scale-invariant curvature fluctuations via the induced entropy fluctuations, it has in fact been shown that the induced non-Gaussianities are larger than the current upper bounds \cite{Lehners}. However, since our mechanism of generating scale-invariant fluctuations is different, the conclusions of \cite{Lehners} will not apply.

\section*{Acknowledgement}

\noindent The research at McGill is supported in part by funds from NSERC and from the Canada Research Chair program. RB is grateful for hospitality of the Institute for Theoretical Physics and the Institute for Particle Physics and Astrophysics of the ETH Zurich during the completion of this project. ZW acknowledges partial support from a McGill Space Institute Graduate Fellowship and from a Templeton Foundation Grant.

\section*{Appendix}

In this Appendix we demonstrate that the S-brane does not effect the equation of motion for the scalar fluctuation variable $u$ discussed in this paper. This contrasts to the large effect which the S-brane has on the evolution of the $v$ variable and which we discussed in our previous paper \cite{us}. 

The starting point of our analysis is the total action (\ref{action}). We insert our ansatz for metric and matter fluctuations (\ref{pertmetric}) and (\ref{pertmatter}) into this action and expand to second order in our metric potential variable $\Phi$. The calculation is explained in detail in \cite{MFB}, and we only quote the crucial points. We need to focus on the relative normalizations with which $\Phi$ enters the bulk and the brane part of the action. The term in the bulk action quadratic in $\Phi^{\prime}$ is (see Eq. (10.35) in \cite{MFB})
\be \label{pertbulk}
S^{(2)} \, = \, \frac{3}{8 \pi G} \int d^4x a^2 \bigl[ - {\Phi^{\prime}}^2 - 4 {\cal{H}} \Phi \Phi^{\prime}
- 6 {\cal{H}}^2 \Phi^2 - (\partial_i \Phi)^2 \bigr] \, .
\ee
On the other hand, the term quadratic iin the brane action is
\be
S_B^{(2)} \, = \,  \frac{1}{2} \int d^4x \kappa a^3 3 \Phi^2 \delta(\tau - \tau_B) \, .
\ee
The resulting equation of motion for $\Phi$ is (in Fourier space)
\be
\Phi^{\prime \prime} + 2 {\cal{H}} \Phi^{\prime}
+ \bigl[k^2 + 2 {\cal{H}}^2 - 2 {\cal{H}}^{\prime} - 4 \pi G a \kappa \delta(\tau - \tau_B) \bigr] \Phi \, = \, 0 \, .
\ee

We see that the brane contribution is suppressed compared to the bulk contribution by a factor ${\cal{F}}$ which is
\be
{\cal{F}} \, \sim \, a \kappa m_{pl}^{-2}  \, .
\ee
Note that ${\cal{F}}$ has mass dimension one since the brane contribution is multiplied by $\delta(\tau - \tau_B)$. The brane thus yields a $\delta$ function contribution to the effective mass in the equation of motion (\ref{ueq}) for $u$ which becomes
\be \label{ueq2}
u_k^{\prime \prime} + \bigl[ k^2 - \frac{2}{a} {\cal{H}}^2 - \frac{a^{\prime \prime}}{a} +
{\cal{F}} \delta(\tau - \tau_B) \bigr] u_k \, = \, 0 \, .
\ee
We see that, unlike in the case of the equation of motion for the canonical variable $v$ studied in \cite{us}, the contribution of the brane inside the equation of motion for $u$ is suppressed by $a$ and therefore negligible. Hence, the $u_k$ modes do not acquire the infrared enhancement which the $v$ modes experience.

\end{document}